\documentclass[12pt,a4paper]{article}
\usepackage{epsfig}
\usepackage{cite}
\usepackage{amsmath}
\usepackage{amssymb}
\usepackage{subfigure}
\def\gev{{\rm GeV}}

\begin{document}
\title{Rare baryonic decays $\Lambda_b\rightarrow \Lambda l^{+}l^{-}$
in the $TTM$ model
 \hspace*{-0.8cm}  }

\author{Yan-Li Wen, Chong-Xing Yue and Jiao Zhang\\
{\small  Department of Physics, Liaoning Normal University, Dalian
116029, China}\thanks{cxyue@lnnu.edu.cn}\\}
\date{}

\maketitle

\begin{abstract}
In the framework of the top triangle moose $(TTM)$ model, we analyze the
 rare decays $\Lambda _{b}\rightarrow \Lambda l^{+}l^{-} (l=e,\mu,\tau)$ by
 using the form factors calculated in full $QCD$. We calculate the contributions
 of the new particles predicted by this model to observables, such as the branching ratios, forward-backward
 asymmetry ($A_{FB}$) and polarizations related  these decay processes. We find that, in wide range of the parameter space, the values of the branching ratios are enhanced by one order of magnitude comparing to the $SM$ predictions.  This model can also produce significant corrections to $A_{FB}$, normal polarization $P_N$ and transversal polarization  $P_T$.

\vspace{1cm}

$PACS$ numbers: 13.30.Ce, 12.60.Cn

\vspace{1cm}

\end{abstract}

\newpage

\noindent{\bf 1. Introduction }\vspace{0.5cm}

The rare decays $\Lambda _{b}\rightarrow \Lambda l^{+}l^{-} (l=e,\mu,\tau)$
which are induced by the flavor-changing neutral current ($FCNC$) can be described by the processes
$b\rightarrow sl^{+}l^{-}$ at the quark level. In the standard model $(SM)$, these $FCNC$ transitions
are forbidden at tree level, while can occur at loop level. New physics models beyond the $SM$ can appear either through new contributions to the Wilson coefficients that enter into the effective Hamiltonian that describes these decays, or through new operators in the effective Hamiltonian which are absent in the $SM$. They can
not only provide very important consistency check of the $SM$, but also are very sensitive
to some  new physics  models beyond the $SM$. Furthermore, unlike the mesonic
decays, the baryonic decays could maintain the helicity structure of the effective
Hamiltonian for the $b\rightarrow s$ transition exactly explains
why one gives more interest to them\cite{1}.

The $SM$ predictions for branching ratios of the semileptonic decays $\Lambda _{b}\rightarrow \Lambda l^{+}l^{-}$
 have been studied in Ref.\cite{2,2a}, which uses the related form factors
calculated via light cone $QCD$ sum rules in full theory. Their results show
$Br(\Lambda _{b}\rightarrow \Lambda e^{+}e^{-})=(4.6\pm1.6)\times10^{-6}$,
$Br(\Lambda _{b}\rightarrow \Lambda \mu^{+}\mu^{-})=(4.0\pm1.2)\times10^{-6}$,
 and $Br(\Lambda _{b}\rightarrow \Lambda \tau^{+}\tau^{-})=(0.8\pm0.3)\times10^{-6}$.
 The first experimental result in investigation of the rare baryonic decays has recently
 been reported by the $CDF$ collaboration at Fermilab, and they announced the result
 of the branching ratio $Br(\Lambda _{b}\rightarrow \Lambda \mu^{+}\mu^{-})=[1.73\pm0.42(stat)\pm0.55(syst)]
\times10^{-6}$\cite{3}. The $LHCb$ collaboration at $CERN$ has also started
 to search these decay channels. So studying of these rare baryonic decays is now entering a
new interesting era.

As has already been noted, rare decays induced by $b\rightarrow s$ transition
are quite promising for searching new physics beyond the $SM$. In recently years,
many works about decays $\Lambda_b\rightarrow \Lambda l^{+}l^{-}$ have been done in
many new physics models, such as standard model with fourth
 generation ($SM4$)\cite{4,4a,4b}, supersymmetry ($SUSY$) model\cite{5},
 the universal extra dimension$(UED)$ model\cite{6,6a,6b,6c},
two Higgs doublet model $(2HDM)$\cite{7}, family non-universal $Z'$ model
\cite{8,8a,8b} and the covariant constituent quark model\cite{8c}. They have shown that some new physics models beyond the $SM$ can indeed give significant contributions to
the rare decays $\Lambda_b\rightarrow \Lambda l^{+}l^{-}$ and the present or future experimental results can be used to test or restrict these new physics models.

  The large mass of the top quark might has a different origin from masses of other light quarks and leptons,
a top quark condensate, $\langle t\bar{t}\rangle$, could be responsible for at least part of electroweak symmetry breaking $(EWSB)$\cite{9}. The top triangle moose $(TTM)$ model\cite{10, 11} is one of interesting new physics models with separate sectors for dynamically generating the masses of the top quark and
the weak gauge bosons $W^{\pm}$ and $Z$. $EWSB$ results largely
from the Higgsless mechanism while the top quark mass is mainly
generated by the topcolor mechanism. So, in this model, there are two sets of
Goldstone bosons. One set is eaten by the  gauge bosons
$W^{\pm}$, $Z$, $W'^{\pm}$ and $Z'$ to generate their masses, while the other set remans in
the spectrum, which is called the top-pions ($\pi_{t}^{0}$  and
$\pi^{\pm}_{t}$) and the top-Higgs
$h_{t}^{0}$. The properties of these new scalars have been recently  studied in Refs.\cite{11, 12, 13}.  In this paper, we will consider the contributions of the  $TTM$ model to the rare decays $\Lambda _{b}\rightarrow \Lambda l^{+}l^{-} (l=e,\mu,\tau)$ and compare our numerical results with those obtained in the $SM$.

The layout of the present paper is as follows. In section 2, we simply review the
essential features of the $TTM$ model. The contributions of the $TTM$ model to  the observables, such as branching ratios, forward-backward
 asymmetry ($A_{FB}$) and polarizations, which are related the decays $\Lambda_b\rightarrow \Lambda l^{+}l^{-}$,
 are given in section 3. In this section we also  compare our numerical results with predictions of the $SM$. Our conclusion is given in section 4.

\vspace{0.5cm} \noindent{\bf 2. The essential features of the $TTM$
model }

\vspace{0.5cm}The detailed description of the $TTM $ model can be
found in Refs.[17,18], and here we just want to briefly review its
essential features, which are related to our calculation.

The electroweak gauge structure of the $TTM$ model is
$SU(2)_{0}\times SU(2)_{1}\times U(1)_{2}$. The nonlinear sigma
field $\sum_{01}$ breaks the group $SU(2)_{0}\times SU(2)_{1}$ down to
$SU(2)$ and field $\sum_{12}$ breaks $SU(2)_{1}\times U(1)_{2}$
down to $U(1)$. To separate top quark mass generation from $EWSB$, a
top-Higgs field $\Phi$ is introduced to the $TTM$ model, which
couples preferentially to the top quark. To ensure that most of the $EWSB$
comes from the Higgsless side, the $VEVs$ of the fields $\sum_{01}$
and $\sum_{12}$ are chosen to be $<\sum_{01}>=<\sum_{12}>=F=\sqrt2
\nu\cos\omega$, in which $\nu=246 GeV$ is the electroweak scale and
$\omega$ is a new small parameter. The $VEV$ of the top-Higgs field
is $f=<\Phi>=\nu \sin\omega$.

From above discussions, we can see that, for the $TTM$ model, there
are six scalar degrees of freedom on the Higgsless sector and four
on the top-Higgs sector. Six of these Goldstone bosons are eaten to
give masses to the gauge bosons $W^{\pm}$, $Z$, $W'^{\pm}$ and $Z'$.
Others remain as physical states in the spectrum, which are called
the top-pions ($\pi_{t}^{\pm}$ and $\pi_{t}^{0}$) and the top-Higgs
$h_{t}^{0}$. In this paper, we will focus our attention on
the contributions of these new particles to the rare decays $\Lambda _{b}\rightarrow \Lambda l^{+}l^{-}  (l=e,\mu,\tau)$.

In general the couplings of the top-pions and top-Higgs to fermions are model dependent, which depend on the individual left-handed and right-handed rotations in the separate up- and down-quark sectors. According the assumptions given by Ref.\cite{13},
the couplings of the  top-pions $\pi_{t}^{0}$ and $\pi_{t}^{\pm}$ to ordinary
fermions, which are related our calculation, are given by
\begin{eqnarray}
&&\frac{i}{\nu}\left[m_{t}\cot\omega\bar{t}_{L}t_{R}+m_{b}\cot\omega\bar{b}_{L}b_{R}+m_{l}\tan\omega\bar{l}_{L}l_{R}\right]\pi^{0}_{t}\nonumber\\
&&+\frac{i\sqrt{2}}{\nu}\left[m_{t}V_{tb}\cot\omega\bar{t}_{R}b_{L}+m_{b}V_{tb}\tan\omega\bar{t}_{L}b_{R}+m_{t}V_{ts}\cot\omega\bar{t}_{R}s_{L}\right]\pi^{+}_{t}+h.c..
\end{eqnarray}
Here $V_{ij}$ is the $CKM$ matrix elements. The couplings of the top-Higgs $h_{t}^{0}$ to fermions are similar to those of the neutral top-pion $\pi_{t}^{0}$.

Reference[17] has extensively studied the couplings of the new heavy gauge bosons $W'^{\pm}$ and $Z'$ to other particles and has shown that the couplings of these new gauge bosons to two heavy quarks (light partners) are  proportional to $1/x$ with $x$ being a small parameter. However, their couplings to ordinary quarks (light quarks) are very small. At the ideal fermion delocalization case, the coupling $g^{W'ud}$ equals to zero, while the couplings $g^{Z'uu}$ and $g^{Z'dd}$ are  proportional to $x$, in which $u $ and $d$ are light up-  and down-quarks, respectively. Thus the contributions of the  $TTM$ model to the rare decays $\Lambda _{b}\rightarrow \Lambda l^{+}l^{-}$ are mainly come from the new scalars ($\pi_{t}^{\pm}$, $\pi_{t}^{0}$ and $h_{t}^{0}$). In the succedent section, we will calculate the contributions of  these new scalars to  the observables, such as branching ratios, forward-backward
 asymmetry ($A_{FB}$) and polarizations, which are related to the decays $\Lambda_b\rightarrow \Lambda l^{+}l^{-} (l=e,\mu,\tau)$.

\vspace{0.5cm} \noindent{\bf \large 3. Numerical results}

At the quark level, the baryonic decays $\Lambda_b\rightarrow \Lambda l^{+}l^{-}$ can be
described by the $FCNC$ transitions $b\rightarrow sl^+l^-$, whose
effective Hamiltonian in the $SM$ is written as:
\begin{eqnarray}
\mathcal{H}^{eff}&=&\frac{G_{F}\alpha_{em}V_{tb}V^{\ast}_{ts}}{2\sqrt{2}\pi}\left[C_{9}^{eff}\bar{s}
\gamma_{\mu}(1-\gamma_{5})b\bar{l}\gamma^{\mu}l+C_{10}\bar{s}
\gamma_{\mu}(1-\gamma_{5})b\bar{l}\gamma^{\mu}\gamma_{5}l\right.\nonumber\\
&&\left.-2m_{b}C_{7}^{eff}\frac{1}{q^2}\bar{s}i\sigma_{\mu\nu}q^{\nu}
(1+\gamma_{5})b\bar{l}\gamma^{\mu}l\right],
\end{eqnarray}
where $q$ is the sum of 4 momenta of $l^+$ and $l^-$, $G_F$ is the Fermi constant, $\alpha_{em}$
is the fine structure constant. The Wilson coefficients $C_{7}^{eff}$, $C_{9}^{eff}$ and $C_{10}$
represent different interactions respectively, whose specific expressions can be obtained in Refs.\cite{14,15,16}.
The transition matrix elements for $\Lambda_b\rightarrow \Lambda l^{+}l^{-}$ can be obtained by
sandwiching the effective Hamiltonian between the initial and final baryonic states, which can be parameterized in terms of twelve form factors $f_{i}, g_{i},
f_{i}^{T}$ and $g_{i}^{T}(i=1,2,3)$ in full $QCD$ theory and can be expressed in the following
manners:
\begin{eqnarray}
\langle\Lambda(p)|\bar{s}\gamma_{\mu}(1-\gamma_{5})b|\Lambda_{b}(p+q)\rangle&=&\bar{u}_{\Lambda}(p)
\left[\gamma_{\mu}f_{1}(q^2)+i\sigma_{\mu\nu}q^{\nu}f_{2}(q^2)+q^{\mu}f_{3}(q^2)\right.\nonumber\\
&&\left.-\gamma_{\mu}\gamma_{5}
g_{1}(q^2)-i\sigma_{\mu\nu}\gamma_{5}q^{\nu}g_{2}(q^2)-q^{\mu}\gamma_{5}g_{3}(q^2)\right]u_{\Lambda_b}(p+q),\nonumber\\
\langle\Lambda(p)|\bar{s}i\sigma_{\mu\nu}q^\nu(1+\gamma_{5})b|\Lambda_{b}(p+q)\rangle&=& \bar{u}_{\Lambda}(p)
\left[\gamma_{\mu}f_{1}^{T}(q^2)+i\sigma_{\mu\nu}q^{\nu}f_{2}^{T}(q^2)+q^{\mu}f_{3}^{T}(q^2)\right.\nonumber\\
&&\left.+\gamma_{\mu}\gamma_{5}
g_{1}^{T}(q^2)+i\sigma_{\mu\nu}\gamma_{5}q^{\nu}g_{2}^{T}(q^2)+q^{\mu}\gamma_{5}g_{3}^{T}(q^2)\right]u_{\Lambda_b}(p+q).\nonumber\\
\end{eqnarray}
The specific expressions of these form factors have been calculated in Ref.\cite{2} in the framework
of full $QCD$ theory. Using above transition matrixes, we can get the angular dependent differential decay
rate of the $\Lambda_b\rightarrow \Lambda l^{+}l^{-}$ decay in the whole physical region ${4m_{l}^2}/m_{\Lambda_{b}}^2\leq \hat{s}\leq(1-\sqrt{r})^2$
which has the following form:
\begin{eqnarray}
\frac{d\Gamma}{d\hat{s}dz}&=&\frac{G_{F}^2\alpha_{em}^2m_{\Lambda_{b}}}{16384\pi^5}|V_{tb}V^{\ast}_{ts}|^2\upsilon\sqrt{\lambda}
\left[\Theta(\hat{s})+\Theta_1(\hat{s})+\Theta_2(\hat{s})z^2\right],
\end{eqnarray}
where $z=\cos\theta$, $\theta$ being the angle between the momenta of $\Lambda_{b}$ and $l^-$ in the center of mass of leptons,
 $\hat{s}=q^2/m_{\Lambda_{b}}^{2}$, $r=m_\Lambda/m_{\Lambda_{b}}$, $\lambda=\lambda(1,r,\hat{s})=
1+r^2+\hat{s}^2-2r-2\hat{s}-2r\hat{s}$ and $\upsilon=\sqrt{1-\frac{4m_l^2}{q^2}}$ is the lepton velocity.
The functions $\Theta(\hat{s}), \Theta_1(\hat{s}), \Theta_2(\hat{s})$ are given by Ref.\cite{2}.

Integrating out the angular dependent differential decay rate, the branching ratios can be obtained as
following manner:
\begin{eqnarray}
Br(\Lambda_b\rightarrow \Lambda l^{+}l^{-})&=&\frac{\tau G_{F}^2\alpha_{em}^2 m_{\Lambda_{b}}|V_{tb}V^{\ast}_{ts}|^2}{8192\pi^5}
\int_{\frac{4m_{l}^{2}}{m^{2}_{\Lambda_{b}}}}^{(1-\sqrt{r})^2}
\upsilon\sqrt{\lambda}\left[\Theta(\hat{s})+\frac{1}{3}\Theta_2(\hat{s})\right]d\hat{s},
\end{eqnarray}
where $\tau $ is the lifetime of $\Lambda_{b}$.

The forward-backward  asymmetry $A_{FB}$ is defined in terms of the differential decay rate as [26]:
\begin{eqnarray}
A_{FB}(\hat{s})=\frac{\int_{0}^{1}\frac{d\Gamma}{d\hat{s}dz}(z,\hat{s})dz-\int_{-1}^{0}\frac{d\Gamma}
{d\hat{s}dz}(z,\hat{s})dz}{\int_{0}^{1}\frac{d\Gamma}{d\hat{s}dz}(z,\hat{s})dz+\int_{-1}^{0}\frac{d\Gamma}{d\hat{s}dz}(z,\hat{s})dz}.
\end{eqnarray}

The normal polarization $P_N$ and transversal polarization  $P_T$ are defined as:
\begin{eqnarray}
P_{i}(q^{2})=\frac{\frac{d\Gamma}{d\hat{s}}(\overrightarrow{\xi}=\overrightarrow{e_{i}})
-\frac{d\Gamma}{d\hat{s}}(\overrightarrow{\xi}=-\overrightarrow{e_{i}})}{\frac{d\Gamma}{d\hat{s}}
(\overrightarrow{\xi}=\overrightarrow{e_{i}})+\frac{d\Gamma}{d\hat{s}}(\overrightarrow{\xi}=-\overrightarrow{e_{i}})},
\end{eqnarray}
where the unit vector $\overrightarrow{\xi}$ represent the spin direction  along $\Lambda$ baryon, $i=N$ or $T$. The explicit expressions of them can be obtained in Ref.\cite{17}.

For the decay processes $\Lambda_b\rightarrow \Lambda l^{+}l^{-}$, the $TTM$ model can give
new contributions to the Wilson coefficients $C_{7}^{eff}$, $C_{9}^{eff}$ and $C_{10}$  by effecting the Inami-Lim functions $C_{0}(x_{t})$,
$D_{0}(x_{t})$, $E_{0}(x_{t})$ and $E'_{0}(x_{t})$ whose explicit expressions can be
obtained in Ref.\cite{18}. In the $TTM$ model,  the detailed expressions of the corresponding functions $C^{TTM}_{0}(x_{t})$,
$D^{TTM}_{0}(x_{t})$, $E^{TTM}_{0}(x_{t})$, $E'^{TTM}_{0}(x_{t})$ including the contributions of the new scalars ($\pi_{t}^{\pm}$, $\pi_{t}^{0}$ and $h_{t}^{0}$) are
shown as:
\begin{eqnarray}
C_{0}^{TTM}(y_{t})&=&\frac{M^2_{\pi}\cot^2\omega}{\sqrt{2}G_{F}M^2_{W}\nu^2}\left(-\frac{y_{t}^2}{8(y_{t}-1)}
-\frac{y_{t}^2}{8(y_{t}-1)^2}\ln[y_{t}]\right),\nonumber\\
D_{0}^{TTM}(y_{t})&=&\frac{\cot^2\omega}{\sqrt{2}G_{F}\nu^2}\left(\frac{47-79y_{t}+38y_{t}^2}{108(1-y_{t})^3}
+\frac{3-6y_{t}^2+4y_{t}^3}{18(1-y_{t})^4}\ln[y_{t}]\right),\nonumber\\
E_{0}^{TTM}(y_{t})&=&\frac{\cot^2\omega}{\sqrt{2}G_{F}\nu^2}\left(\frac{7-29y_{t}
+16y_{t}^2}{36(1-y_{t})^3}-\frac{3y_{t}^2-2y_{t}^3}{6(1-x)^4}\ln[y_{t}]\right),\nonumber\\
E'^{TTM}_{0}(y_{t})&=&\frac{\cot^2\omega}{2\sqrt{2}G_{F}\nu^2}\left(-\frac{5-19y_{t}+
20y_{t}^2}{6(y_{t}-1)^3}+\frac{y_{t}^2-2y_{t}^3}{(1-y_{t})^4}\ln[y_{t}]\right),
\end{eqnarray}
where $y_{t}=m^{2}_{t}/M_{\pi}^{2}$ and we have taken $M_{\pi}=M_{\pi^{0}_{t}}=M_{h^{0}_{t}}=M_{\pi_{t}^{\pm}}$ .
\begin{figure}
\centering
 \subfigure[$$]{
\includegraphics[scale=0.6]{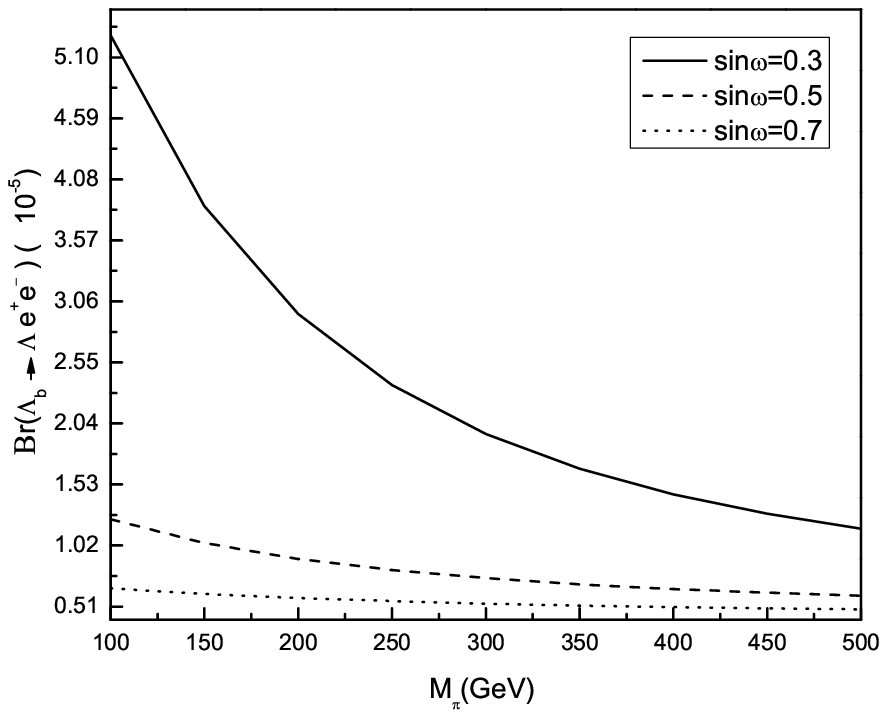}}
\subfigure[$$]{
\includegraphics[scale=0.6]{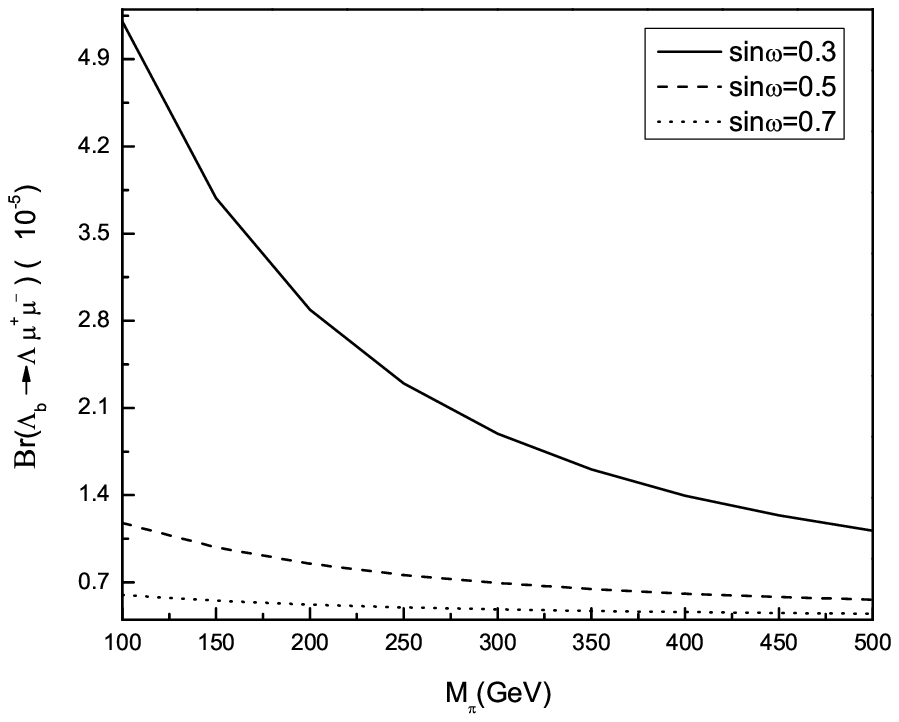}}
\subfigure[$$]{
\includegraphics[scale=0.6]{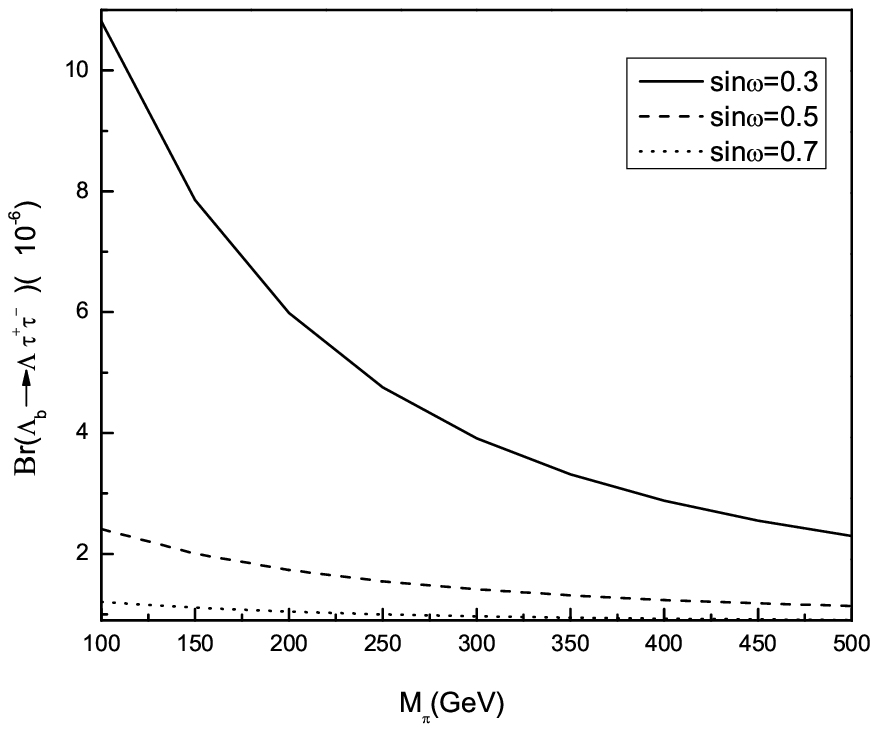}}
\caption{The branching ratios $ Br(\Lambda _{b}\rightarrow \Lambda e^{+}e^{-})$, $Br(\Lambda _{b}\rightarrow \Lambda \mu^{+}\mu^{-})$
and $Br(\Lambda _{b}\rightarrow  \hspace*{1.7cm}\Lambda \tau^{+}\tau^{-})$ as functions of $M_{\pi}$ with  $\sin\omega=0.3, 0.5$, and $ 0.7$ in the $TTM$ model.}
\end{figure}

In our numerical calculation, we take the $SM$ input parameters as\cite{19}:
$M_W= 80.425 \; \gev$, $G_F = 1.166 \times 10^{-5} \; \gev^{-2}$, $\alpha_{em} = 1/129$, $m_{t}=172 \; \gev$,
 $m_{\Lambda_{b}}=5.6202 \; \gev$, $m_\Lambda=1.115683\; \gev$, $\tau_{B_s} = 1.383 \times 10^{-12}s$, $V_{tb}= 0.998$,
 $V_{ts}= 0.042$ and $sin^2\theta_W=0.2312$. The observables about the decay processes
$\Lambda _{b}\rightarrow \Lambda l^{+}l^{-}$ depend on the model dependent parameters:
 the mass of scalars $M_{\pi}$ and the free  parameter $\sin\omega$, which indicates
the fraction of $EWSB$ provided by the top condensate. The top-pion masses depend on the amount of top-quark mass arising
from the $ETC$ sector and on the effects of electroweak gauge interactions, and thus their values  model-dependent. In the context of the $TTM$ model, Ref.[18] has obtained  the constraints on the top-pion mass via studying its effects on the relevant experimental observables. In our numerical calculation, we will assume that the values of the free parameters $\sin\omega$ and $M_{\pi}$ are in the ranges of
$0.2 \sim 0.8$ and $200 \sim 600GeV$, respectively.

Considering the contributions of the $TTM$ model to the rare decays
$\Lambda _{b}\rightarrow \Lambda l^{+}l^{-}$, the branching ratios $Br(\Lambda _{b}\rightarrow \Lambda l^{+}l^{-})$
are plotted in Fig.1 as functions of mass parameter $M_{\pi}$ with the free  parameter $\sin\omega=0.3, 0.5$, and $ 0.7$, in which Fig.1 (a), (b), and (c) represent the results of $Br(\Lambda _{b}\rightarrow \Lambda e^{+}e^{-})$,
$Br(\Lambda _{b}\rightarrow \Lambda \mu^{+}\mu^{-})$ and $Br(\Lambda _{b}\rightarrow \Lambda \tau^{+}\tau^{-})$,
respectively. We can see that enhancing the values of the mass parameter
$M_{\pi}$ or the free parameter  $\sin\omega$ can make the values of the branching ratios decrease. Comparing to the $SM$ predictions,
one easily see that the values of branching ratios $Br(\Lambda _{b}\rightarrow \Lambda l^{+}l^{-})$ can be enhanced by about one order of magnitude in wide range of the parameter space of the $TTM$ model. It is obvious that the experimental measurement value of the branching ratio
$Br(\Lambda _{b}\rightarrow \Lambda \mu^{+}\mu^{-})=[1.73\pm0.42(stat)\pm0.55(syst)]\times10^{-6}$ [3]
is smaller than that given by the $TTM$ model. The more experimental data can give constraints on the  free parameters of the
$TTM$ model in the future.

\begin{figure}
\centering
 \subfigure[$$]{
\includegraphics[scale=0.6]{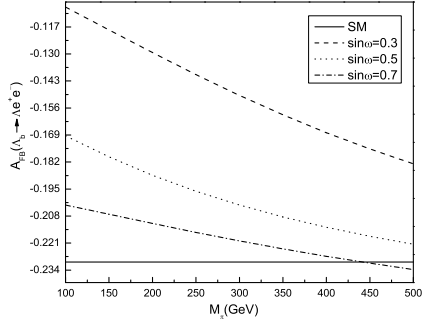}}
\subfigure[$$]{
\includegraphics[scale=0.6]{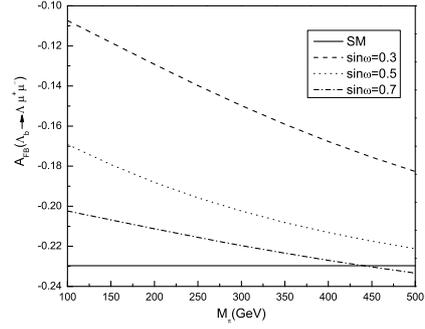}}
\subfigure[$$]{
\includegraphics[scale=0.6]{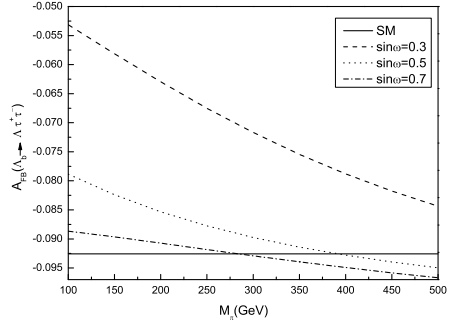}}
\caption{The forward-backward  asymmetry $A_{FB}(\Lambda _{b}\rightarrow \Lambda e^{+}e^{-})$(a), $A_{FB}(\Lambda _{b}\rightarrow \hspace*{1.7cm}\Lambda \mu^{+}\mu^{-})$(b)
and $A_{FB}(\Lambda _{b}\rightarrow \Lambda \tau^{+}\tau^{-})$(c) are plotted as functions of $M_{\pi}$ for \hspace*{1.7cm} different values of the free parameter  $\sin\omega$ in the $TTM$ model.}
\end{figure}

\begin{figure}
\centering
 \subfigure[$$]{
\includegraphics[scale=0.6]{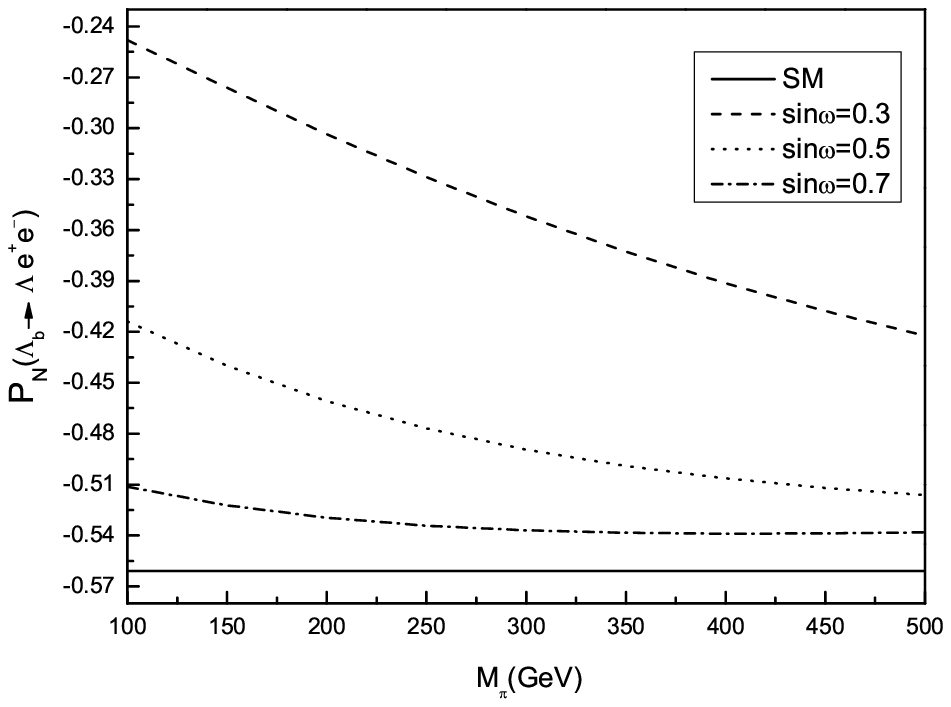}}
\subfigure[$$]{
\includegraphics[scale=0.6]{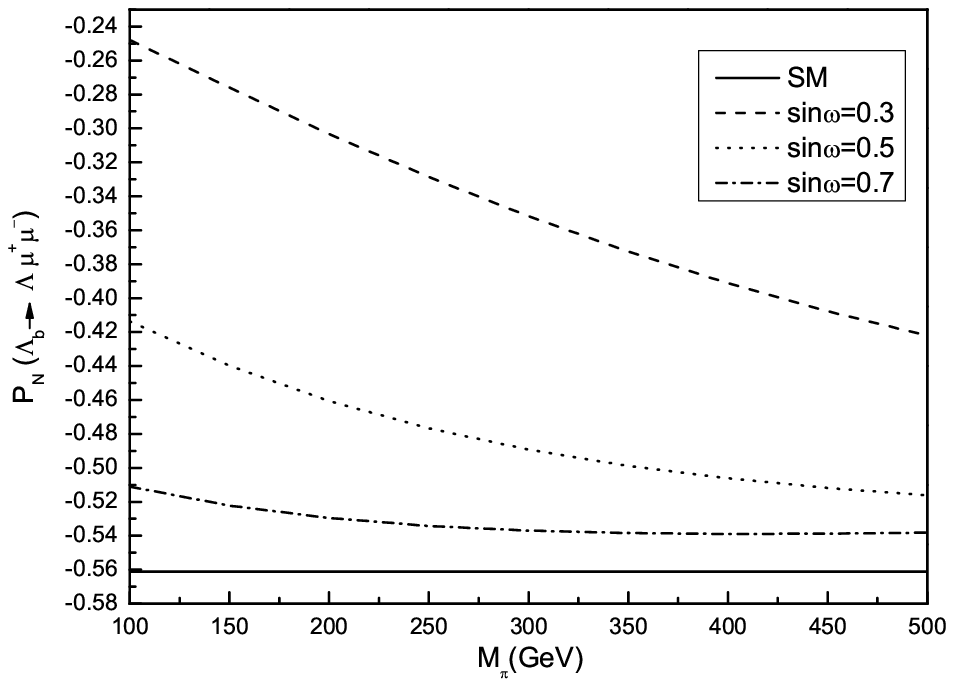}}
\subfigure[$$]{
\includegraphics[scale=0.6]{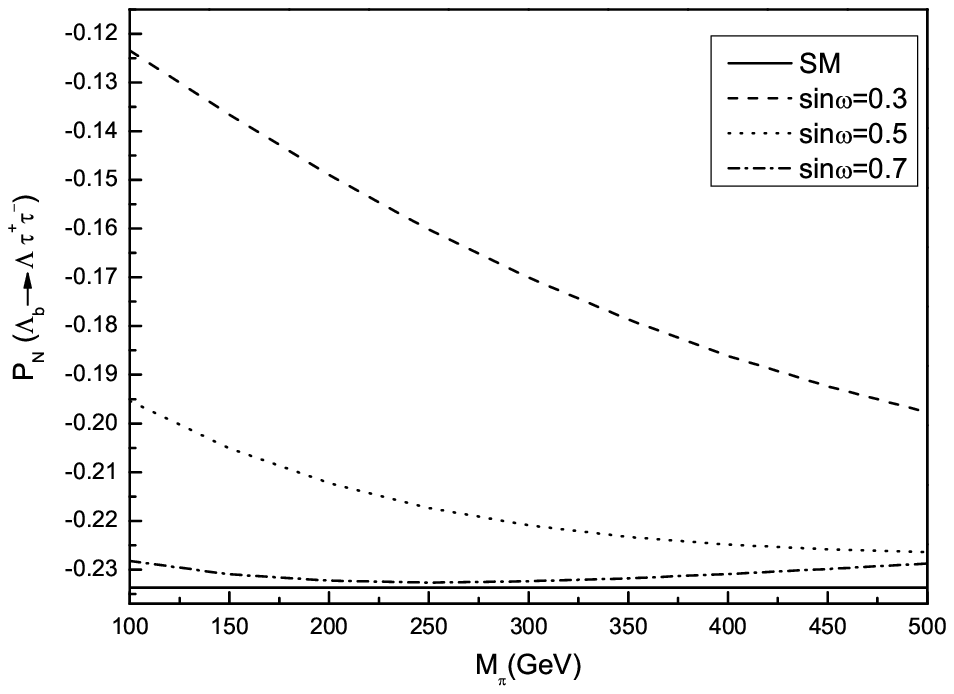}}
\caption{ The normal polarization $P_N(\Lambda _{b}\rightarrow \Lambda l^{+}l^{-})$ as function of $M_{\pi}$ for  different values \hspace*{1.7cm} of the free parameter  $\sin\omega$ in the $TTM$ model.}
\end{figure}

Our numerical results for the forward-backward  asymmetry $A_{FB}(\Lambda _{b}\rightarrow \Lambda l^{+}l^{-})$ are given in Fig.2, in which the horizontal solid line represent their $SM$ predictions.
 From these figures, we can see that  the absolute values of $A_{FB}$ increase as the
increasing of the free parameters $M_{\pi}$ and $\sin\omega$. In wide range of the parameter space, the $TTM$ model can produce positive contributions to the forward-backward  asymmetry $A_{FB}$ and the absolute values of
$A_{FB}$ are smaller  than the corresponding $SM$ predictions.

\begin{figure}
\centering
 \subfigure[$$]{
\includegraphics[scale=0.6]{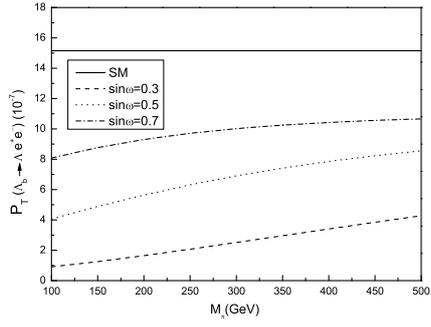}}
\subfigure[$$]{
\includegraphics[scale=0.6]{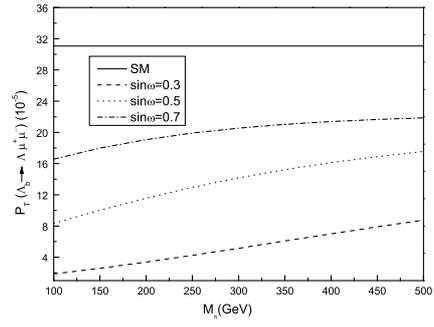}}
\subfigure[$$]{
\includegraphics[scale=0.6]{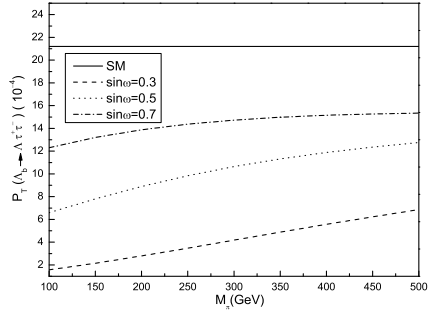}}
\caption{ The transversal polarization  $P_T(\Lambda _{b}\rightarrow \Lambda l^{+}l^{-})$ as function of $M_{\pi}$ for  different \hspace*{1.7cm} values  of the free parameter  $\sin\omega$ in the $TTM$ model.}
\end{figure}

The numerical results for the normal polarization $P_N(\Lambda _{b}\rightarrow \Lambda l^{+}l^{-})$ and transversal polarization  $P_T(\Lambda _{b}\rightarrow \Lambda l^{+}l^{-})$ in the $TTM$ model are given in Fig.3 and Fig.4, respectively, in which the horizontal solid line represent their $SM$ predictions.
It is obvious that the absolute values of $P_{N}(\Lambda _{b}\rightarrow \Lambda l^{+}l^{-})$
increase as the increasing of free parameters $M_{\pi}$ and $\sin\omega$, while the values of $P_{T}(\Lambda _{b}\rightarrow \Lambda l^{+}l^{-})$
increase as these free parameters increasing.  In most of the parameter space of the $TTM$, the absolute values of $P_{N}$ are smaller than those for the $SM$ predictions, but the values of $P_{T}$ are much smaller than  the corresponding $SM$ predictions. However, for large values for the free parameters $M_{\pi}$ and $\sin\omega$, all values of $P_{N}$ and $P_{T}$ approach the values of the $SM$ predictions. This means that the contributions of the $TTM$ model to polarization observables become smaller for large values of the relevant free parameters.

\vspace{0.5cm} \noindent{\bf \large 4. Conclusions}

It is well known that $FCNC$ transitions $b\rightarrow s$ are considered as excellent probes of new physics models beyond
the $SM$. Combing Higgsless and topcolor mechanisms, a new physics model was proposed, called the $TTM$ model, which can be seen as the
deconstructed version of the topcolor-assisted technicolor ($TC2$)
model. This model predicts the new gauge bosons and scalars, which can produce significant contributions to some observables.  We consider the decay processes $\Lambda _{b}\rightarrow \Lambda l^{+}l^{-}$ in the context of this model.

We have calculated the contributions of the $TTM$ model to the branching ratios, forward-backward
asymmetry and polarizations related the decay channels $\Lambda _{b}\rightarrow \Lambda l^{+}l^{-}$ using
the form factors obtained from full $QCD$. The numerical results indicate that, due to the small couplings of the new heavy gauge bosons $W'^{\pm}$ and $Z'$  with the $SM$ fermions, their contributions can be safely neglected and  the contributions of the $TTM$ model to observables mainly come from the new scalars ($\pi_{t}^{\pm}$, $\pi_{t}^{0}$ and $h_{t}^{0}$). In wide range of the parameter space, its contributions to branching ratios $Br(\Lambda _{b}\rightarrow \Lambda l^{+}l^{-})$ can enhance  the corresponding $SM$ predictions by about one order of magnitude. In most of the parameter space of the $TTM$, their values are larger than those in the $SM4$ theory [4,5,6] or in the $SUSY$ model [7]. The $TTM$ model can also produce significant corrections to the observables  $A_{FB}$, $P_{N}$ and $P_{T}$, while their values are larger or smaller than those given by the $SM4$ theory or the $SUSY$ model depending on the relevant free parameters. Certainly, the errors of the form factors [2] can make our numerical results has uncertainties. However, the theoretical uncertainties are much smaller the discrepancies between the $TTM$ and $SM$ predictions. Thus, we expect that our results will be helpful to constrain or test the $TTM$ model at the $LHCb$ via the decay processes $\Lambda _{b}\rightarrow \Lambda l^{+}l^{-}$.

\section*{Acknowledgments} \hspace{5mm}This work was
supported in part by the National Natural Science Foundation of
China under Grants Nos.10975067 and 11275088, the Natural Science Foundation of the Liaoning Scientific Committee
(No. 201102114), and Foundation of Liaoning Educational Committee (No. LT2011015).

\end{document}